\documentclass[conference]{IEEEtran}
\IEEEoverridecommandlockouts
\usepackage{amsmath, amssymb, cite, tikz, graphicx, xspace, mathrsfs, mathtools, psfrag, chemarrow, subfigure,color,soul,graphicx,mathtools}
\usepackage{kbordermatrix}
\usepackage{enumitem}
\usepackage{bbm}
\usepackage[utf8]{inputenc}
\usepackage{booktabs}
\usepackage{amscd}
\usepackage{amsmath}
\usepackage{amssymb}
\usepackage{amsthm}

\usepackage{epsfig}
\usepackage{verbatim}
\usepackage{graphicx}
\usepackage{amsthm}
\usepackage{color}
\usepackage{ multirow }
\usepackage{algorithm}
\usepackage[noend]{algpseudocode}
\usetikzlibrary{patterns}
\usetikzlibrary{angles} 
\usepackage{tikz}
\usetikzlibrary{shapes, arrows, decorations.markings, arrows.meta}
\usetikzlibrary{patterns} 
\usetikzlibrary{arrows,automata,positioning,chains,calc}
\usetikzlibrary{quotes,angles}
\newtheorem*{remark}{Remark}
\makeatletter
\def\BState{\State\hskip-\ALG@thistlm}
\makeatother
\def\BibTeX{{\rm B\kern-.05em{\sc i\kern-.025em b}\kern-.08em
    T\kern-.1667em\lower.7ex\hbox{E}\kern-.125emX}}
\begin{document}
\title{Real-time Remote Reconstruction of a Markov Source and Actuation over Wireless}
\author{
	\IEEEauthorblockN{Mehrdad Salimnejad\IEEEauthorrefmark{1}, 
	Marios Kountouris\IEEEauthorrefmark{2}, and Nikolaos Pappas\IEEEauthorrefmark{1}}
    \IEEEauthorblockA{\IEEEauthorrefmark{1}Department of Computer and Information Science, Link\"{o}ping University, Link\"{o}ping, Sweden}
    \IEEEauthorblockA{\IEEEauthorrefmark{2}Communication Systems Department, EURECOM, Sophia-Antipolis, France}
   \{mehrdad.salimnejad, nikolaos.pappas\}@liu.se, marios.kountouris@eurecom.fr
}

\maketitle
\begin{abstract}
In this work, we study the problem of real-time tracking and reconstruction of an information source with the purpose of actuation. A device monitors an $N$-state Markov process and transmits status updates to a receiver over a wireless erasure channel. We consider a set of joint sampling and transmission policies, including a semantics-aware one, and we study their performance with respect to relevant metrics. Specifically, we investigate the real-time reconstruction error and its variance, the consecutive error, the cost of memory error, and the cost of actuation error. Furthermore, we propose a randomized stationary sampling and transmission policy and derive closed-form expressions for all aforementioned metrics. We then formulate an optimization problem for minimizing the real-time reconstruction error subject to a sampling cost constraint. 
Our results show that in the scenario of constrained sampling generation, the optimal randomized stationary policy outperforms all other sampling policies when the source is rapidly evolving.
Otherwise, the semantics-aware policy performs the best.
\end{abstract}
\vspace{-0.05in}
\section{Introduction}
Networked control systems have recently received significant attention due to their promise of enabling various use cases, such as swarm robotics for target tracking, healthcare systems, autonomous transportation and environmental surveillance using sensor networks, to name a few. A key functionality in these systems entails a device sending time-stamped status updates to a remote monitor, which is tasked to track the state of the monitoring process. Therefore, a relevant yet highly challenging problem in this context is that of real-time remote tracking and of deriving a sampling policy that minimizes error performance metrics of the reconstructed process.

Several studies have been carried out in this area. Different variants of remote state estimation under communication constraints for linear time-invariant (LTI) systems are considered in \cite{lipsa2011remote,shi2012scheduling,wu2018optimal,nayyar2013optimal}. Fundamental limits and trade-offs of remote estimation of Markov processes in real-time communication are studied in \cite{chakravorty2015distortion,chakravorty2016fundamental}. The problem of scheduling in event triggered estimation is considered in \cite{leong2016sensor}. 
Optimal sampling of stochastic processes for minimizing the mean square estimation error is studied in \cite{sun2019sampling,ornee2021sampling,GuoKostina2022}.
In \cite{hui2022real}, a joint sampling and quantization policy is presented for real-time monitoring of a Brownian motion. The problem of whether to retransmit a previous sample or a new sample in real-time remote estimation is studied in \cite{huang2019retransmit}.
More recently, new metrics that account for the importance and the effectiveness (semantics) of information \cite{ShannonWeaver49,VoI_USSR,QoI} with respect to the goal of data exchange have been introduced among others in \cite{kountouris2021semantics,Qin22arxiv,tolga21SP,PetarProc2022,kalfa2022reliable,GunduzJSAC23}. 
The most closely related work is \cite{pappas2021goal}, where new goal-oriented semantic sampling and communication policies are proposed for the problem of real-time tracking and source reconstruction of a two-state Markov process. 

In this paper, we extend \cite{pappas2021goal} in several aspects. First, we employ a general $N$-state discrete time Markov chain (DTMC) model to describe the information source/process evolution, as depicted in Fig. \ref{DTMC_NStates}. Second, we consider joint sampling and transmission and propose a set of metrics for system performance evaluation. We derive general expressions for the real-time reconstruction error and we introduce two new timing-aware error metrics, namely the consecutive error and the cost of memory error, which capture the effect on the performance when the system remains in an erroneous state for several consecutive time slots.
Furthermore, we consider the cost of actuation error to investigate the significance or the non-commutative effects of an error at the receiver side since different errors may have different impact on the system. Then, we formulate and solve a constrained optimization problem where the objective function is the time-averaged reconstruction error subject to a sampling cost constraint. Comparing different sampling and transmission policies, we show under which conditions the semantics-aware policy outperforms the rest and when the proposed randomized stationary policy can be beneficial.

\vspace{-0.5cm}
\section{System Model}
\label{system model}
\par
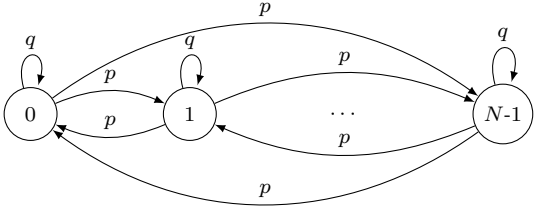
\begin{figure}[!t]
	\centering
	\begin{tikzpicture}[start chain=going left,->,>=latex,node distance=1.4cm]
	\footnotesize
		\node[state, on chain]        
		(N-1){$N$-$1$};
		\node[on chain]                        (g) {$\cdots$};
		\node[state, on chain]                 (2){$1$};
		\node[state, on chain]                 (1){$0$};
		\draw[>=latex]
		
		(1)   edge[loop above] node {$q$} ()
		(1)   edge[bend left=23] node[above] {$p$}(2)
		(1)   edge[bend left=36] node[above]{$p$} (N-1)
		(2)   edge[loop above] node {$q$} ()
		(2)   edge[bend left=23] node[above] {$p$}(1)
		(2)   edge[bend left=23] node[above]{$p$} (N-1)
		(N-1)   edge[loop above] node {$q$} ()
		(N-1)   edge[bend left=23] node[above] {$p$}(2)
		(N-1)   edge[bend left=36] node[above] {$p$}(1)
		;
	\end{tikzpicture}
	\setlength\abovecaptionskip{-0.7\baselineskip}
	\caption{DTMC describing the evolution of the information source.}
	\label{DTMC_NStates}
\end{figure}

 We consider a time-slotted communication system in which a sampler performs sampling of a process $X_{t}$ at time slot $t$, and then the transmitter informs the receiver by sending samples over a wireless communication channel. We model the information source by an $N$-state discrete time Markov chain (DTMC) $\{X_t, t \in \mathbb{N}\}$, depicted in Fig.\ref{DTMC_NStates}. Therein, the self-transition probability and the probability of transition to another state at time slot $t+1$ are defined as $\text{Pr}\big[X(t+1) = X(t)\big] = q$ and $\text{Pr}\big[X(t+1) \neq X(t)\big] = p$, respectively. Since the process of $X_t$ can have one of $N$ different possible values, we can write $q+(N-1)p = 1$.
In this paper, we denote the action of sampling at time slot $t$ by $\alpha^{\text{s}}_{t}$, where $\alpha^{\text{s}}_{t}=1$ if the source is sampled and $\alpha^{\text{s}}_{t} = 0$ otherwise. The action of transmitting a sample is defined as $\alpha^{\text{tx}}_{t}$, where $\alpha^{\text{tx}}_{t}=1$ if a sample is transmitted, otherwise the transmitter remains idle, i.e., $\alpha^{\text{tx}}_{t}=0$.

\subsection{Physical Layer Model}
We assume that the communication channel between transmitter and receiver is subject to small-scale Rayleigh fading and large-scale pathloss attenuation. The received power is given by ${P}_{\text{rx}} = {P}_{\text{tx}}g r^{-\beta}$
where $P_{\text{tx}}$ is the transmitted power, and $g$ is fading power between transmitter and receiver. We assume that $g$ is independent and identically distributed (i.i.d) random variable (RV) with unit mean and cumulative distribution function (CDF) $F_{g}(x) = 1-e^{-x}, x \geqslant 0$. The distance between transmitter and receiver is denoted by $r$, where $\beta > 2$ is the pathloss exponent. At each time slot, the received signal to noise ratio (SNR) is given by $\text{SNR} = \frac{P_{\text{rx}}}{\sigma^2},$
where $\sigma^2$ is the variance of the complex additive white Gaussian noise (AWGN) at the receiver. At time slot $t$, the receiver constructs an estimate of the process $X_{t}$, denoted by $\hat{X}_{t}$. The channel state $h_{t}$ is equal to $1$ if a sample is successfully decoded by the receiver and $0$ otherwise. It is assumed that a sample is successfully received if the received SNR exceeds a certain threshold. The probability that a sample is successfully decoded is given by
\begin{align}
	\label{success_Prob}
	p_{s} = \mathrm{Pr}\big[h_{t}=1\big]= \mathrm{Pr}\big[\text{SNR}>\gamma\big] = \exp\left(-\frac{\gamma\sigma^{2}}{P_{\text{tx}}r^{-\beta}}\right)
\end{align}
where $\gamma$ is an SNR threshold. We consider that a successful transmission is declared to the transmitter using an acknowledgment (ACK) packet. The receiver also sends a negative-ACK packet in the case of a transmission failure. It is assumed that ACK/NACK packets are delivered instantaneously and error free to the transmitter. Therefore, the transmitter has perfect knowledge of the reconstructed source state at time slot $t$, i.e., $\hat{X}_{t}$. We also assume that a sample is discarded when its transmission fails.

\subsection{Sampling and Transmission Policies}
\label{SamplingPolicies}
\par We introduce here the four sampling and transmission policies considered in this work.
\begin{enumerate}
    \item \textbf{Uniform Policy:} sampling is performed periodically every $d$ time slot, independently of the evolution of the source $X_{t}$. Therefore, the sampling time sequences are $\{t_{k} = k d, k\geqslant 1\}$. Although this policy's implementation is simple, several state transitions can be missed between two consecutive sampling events. 
    \item \textbf{Change-Aware Policy:} the generation of a new sample is triggered when a change at the state of the source $X_{t}$ between two consecutive time slots is observed, regardless of whether the system is in synced state or not. 
    \item \textbf{Semantics-Aware Policy:} sample generation is triggered in two cases. First, in the case where at a given time slot $t$ the system is in synced state, i.e., $X_{t}=\hat{X}_{t}$, sampling is performed if a change at the state of the source at time slot $t+1$ occurs, i.e., $X_{t+1}\neq {X}_{t}$. Second, in the case where at time slot $t$, the system is in an erroneous state, i.e., $X_{t}\neq \hat{X}_{t}$, sample acquisition is triggered whenever $X_{t+1}\neq \hat{X}_{t}$. 
    \item \textbf{Randomized Stationary Policy:} sampling is performed in a probabilistic manner at each time slot.
\end{enumerate}
\begin{remark}
The uniform, change-aware, and randomized stationary policies do not require an ACK/NACK feedback channel from the receiver to the transmitter.
\end{remark}
\section{Real-time Reconstruction Error}
\label{RealTime}
\par The real-time reconstruction error is defined as the difference between $X_{t}$ and $\hat{X}_{t}$ at time slot $t$, i.e.,
\begin{align}
	\label{reconstructed_error}
	E_{t} = \left|X_{t}- \hat{X}_{t}\right|.
\end{align}
The time-averaged reconstruction error for an observation interval $[1, T]$, with $T$ being a large positive number, is defined as
\begin{align}
	\label{timeAvg_Error}
	\bar{E} = \lim_{T\to\infty} \frac{1}{T}\sum_{t = 1}^{T} \mathbbm{1}\left(E_{t}\neq 0\right) = \lim_{T\to\infty} \frac{1}{T}\sum_{t = 1}^{T} \mathbbm{1}\left(X_{t}\neq \hat{X}_{t}\right),
\end{align}
where $\mathbbm{1}(\cdot)$ is the indicator function. We define the transition probabilities of $E_{t}$ as
\begin{align}
	\label{trans_prob}
	P_{i,j} = \mathrm{Pr}\big[E_{t+1} \!=\! j\big|E_{t} = i\big],\hspace{0.05 cm} \forall i,j\in\{0,1,\cdots, N-1\}
\end{align}
where at time slot $t$ the synced state of the system is denoted by $E_{t} = 0$, while $E_{t}\neq 0$ denotes the system is in an erroneous state. 
In the following subsections, we derive general expressions for the transition probabilities of $E_{t}$ under randomized stationary policy for an $N$-state DTMC information source.

\subsection{Transition Probabilities}
\label{NState1}
Using the total probability theorem, one can write the transition probabilities $P_{i,j}$, given in \eqref{trans_prob}, as
\begin{align}
	\label{trans_prob2}
	&P_{i,j} \notag\\
	&\!\!=\!\!\! \sum_{n=0}^{N-1} \mathrm{Pr}\big[E_{t+1} = j\big|E_{t} = i, X_{t} = n\big]\mathrm{Pr}\big[X_{t} = n\big|E_{t} = i\big]\notag\\
	&\!\!=\!\!\!\sum_{n=0}^{N-1} \frac{\mathrm{Pr}\big[E_{t+1} = j\big|E_{t} = i, X_{t} = n\big]\mathrm{Pr}\big[X_{t} = n, E_{t} = i\big]}{\mathrm{Pr}\big[E_{t}=i\big]}\notag\\
	&\!\!=\!\!\!\!\sum_{n=0}^{N-1} \!\!\Bigg(\!\!\frac{c_{1}\!\times\!\!\mathrm{Pr}\big[E_{t+1} \!\!=\!\! j\big|X_{t}\! =\! n,\! \hat{X}_{t}\!=\!n\!-\!i\big]\!\mathrm{Pr}\big[X_{t} \!\!= \!n, \hat{X}_{t}\!=\!n\!\!-\!\!i\big]}{\mathrm{Pr}\big[E_{t}=i\big]},\notag\\
	&+ \frac{c_{2}\!\times\!\mathrm{Pr}\big[E_{t+1} \!\!=\!\! j\big|X_{t}\! =\! n,\! \hat{X}_{t}\!=\!n\!+\!i\big]\!\mathrm{Pr}\big[X_{t} \!\!= \!n, \hat{X}_{t}\!=\!n\!\!+\!\!i\big]}{\mathrm{Pr}\big[E_{t}=i\big]}\Bigg)\notag\\&\hspace{3.5 cm} \forall i,j\in\{0,1,\cdots, N-1\},
\end{align}
where $c_{1}$ and $c_{2}$ in \eqref{trans_prob2} are given by
\begin{align}
	\label{c1_c2}
	c_{1}&= \mathbbm{1}(i=0)0.5+\mathbbm{1}( n\geqslant i, i\neq 0),\notag\\
	c_{2}&=\mathbbm{1}(i=0)0.5+\mathbbm{1}(n+i\leqslant N-1, i\neq 0).
\end{align}
Furthermore, the probability $\mathrm{Pr}\big[E_{t}=i\big]$ in \eqref{trans_prob2} is given by
\begin{multline}
	\label{PEt_i}
	\!\!\!\!\!\!\mathrm{Pr}\big[E_{t}=i\big] = \sum_{m=0}^{N-1} \mathrm{Pr}\big[ X_{t} = m, E_{t}=i\big]\\
	\hspace{0.4cm}=\sum_{m=0}^{N-1}\Bigg(f_{1}\times\mathrm{Pr}\big[X_{t} \!= \!m, \hat{X}_{t} \!= \! m-i\big]+\\
	f_{2}\times\mathrm{Pr}\big[X_{t} \!= \!m, \hat{X}_{t} \!= \! m+i\big]\Bigg) 
\end{multline}
\vspace{1ex}
where $f_{1}$ and $f_{2}$ in \eqref{PEt_i} are given by
\begin{align}
	\label{f1_f2}
	f_{1}&= \mathbbm{1}(i=0)0.5+\mathbbm{1}( m\geqslant i, i\neq 0),\notag\\
	f_{2}&=\mathbbm{1}(i=0)0.5+\mathbbm{1}(m+i\leqslant N-1, i\neq 0).
\end{align}
Note that the joint probability $\mathrm{Pr}\big[X_{t} \!= \! m, \hat{X}_{t} \!\!=\! m\pm i\big]$ is the stationary distribution of the two-dimensional DTMC describing the joint status of the system regarding the current state at the original source. Now, 
the expression $\displaystyle\frac{\mathrm{Pr}\big[X_{t} = n, \hat{X}_{t} =n\pm i\big]}{\mathrm{Pr}\big[E_{t} = i\big]}$ given in \eqref{trans_prob2} can be simplified as
\begin{align}
	\label{Cond_Prob_X_E}
	\!\!\!\frac{\mathrm{Pr}\big[X_{t} \!=\! n, \hat{X}_{t} \!=\!n\pm i\big]}{\mathrm{Pr}\big[E_{t} = i\big]} \!=\! \mathbbm{1}(i\!=\!0)\frac{1}{N}\!+\!\mathbbm{1}(i\neq 0)\frac{1}{2(N\!-\!i)}.
\end{align}
To derive the conditional probability given in \eqref{trans_prob2}, we first define ${H}_{0}$ and ${H}_{1}$ as

\begin{align}
	\label{H0_H1}
	{H}_{0} &= \mathrm{Pr}\Big[\alpha^{\text{s}}_{t+1}=1, \alpha^{\text{tx}}_{t+1}=1, h_{t+1} = 0\Big]\notag\\
	&=\mathrm{Pr}\Big[\alpha^{\text{s}}_{t+1}=1, \alpha^{\text{tx}}_{t+1}=1\Big]\mathrm{Pr}\Big[h_{t+1} = 0\Big]\notag\\
	{H}_{1} &= \mathrm{Pr}\Big[\alpha^{\text{s}}_{t+1}=1, \alpha^{\text{tx}}_{t+1}=1, h_{t+1} = 1\Big]\notag\\
	&=\mathrm{Pr}\Big[\alpha^{\text{s}}_{t+1}=1, \alpha^{\text{tx}}_{t+1}=1\Big]\mathrm{Pr}\Big[h_{t+1} = 1\Big],
\end{align}
where ${H}_{0}$ and ${H}_{1}$ are the joint probability density function (PDF) of sampling and transmissions actions at time slot $t+1$ when we have failed and successful transmission, respectively. We define $\mathrm{Pr}\{\alpha^{\text{s}}_{t+1} = 0\}$ as the probability that the source is not sampled at time slot $t+1$, which is equal to $1-\mathrm{Pr}\{\alpha^{\text{s}}_{t+1} = 1\}$. In this work, we assume that the transmitter can send samples immediately after the sampler performs sampling. This means that the joint probability of sampling and transmissions actions at time slot $t+1$ when the sampler performs sampling is equal to $p_{\alpha^{\text{s}}} = \mathrm{Pr}\big[\alpha^{\text{s}}_{t+1}=1\big]$. Therefore, \eqref{H0_H1} can be simplified as
\begin{equation}
	\label{H0_H1_2}
	{H}_{0} = p_{\alpha^{\text{s}}}\big(1-p_{s}\big),
	{H}_{1} = p_{\alpha^{\text{s}}}p_{s},
\end{equation}
where $p_{s}$ is the success probability given in \eqref{success_Prob}. 
Using \eqref{trans_prob2}, \eqref{c1_c2}, and \eqref{Cond_Prob_X_E}, we can derive $P_{0,0}$ as
\begin{align}
	\label{P00}
	P_{0,0} &\!=\!\!\sum_{n=0}^{N-1}\! \frac{\mathrm{Pr}\big[E_{t+1} = 0\big|X_{t} = n, \hat{X}_{t} \!=\! n\big]\mathrm{Pr}\big[X_{t} \!=\! n, \hat{X}_{t} \!=\! n\big]}{\mathrm{Pr}\big[E_{t}=0\big]}\notag\\
	&=\!\sum_{n=0}^{N-1} \frac{1}{N}\mathrm{Pr}\big[E_{t+1} = 0\big|X_{t} = n, \hat{X}_{t} = n\big].
\end{align}
We now calculate the conditional probability in \eqref{P00}. To this end, we first note that the receiver has perfect knowledge of the process $X_{t}$ at time slot $t$. Therefore, the system will be in synced state at time slot $t+1$, i.e., $E_{t+1} \!=\! 0$, if the state of the process $X_{t}$ does not change, which happens with probability $q$. In addition, $E_{t+1}\!=\!0$ when the state of the source changes to one of the remaining $N\!-\!1$ states and the sample is successfully decoded by the receiver. This event occurs with probability $(N-1)p{H}_{1}$. Then \eqref{P00} can be written as
\begin{equation}
	\label{P00_1}
	P_{0,0} =  q+(N\!-\!1)p{H}_{1}.
\end{equation}
Similarly, one can obtain the transition probabilities $P_{i,j}$ for different values of $i$ and $j$ as follows
\vspace{0in}
\begin{align}
	\label{pij_M1_1}
	P_{i,0}&=\mathbbm{1}( 1\leqslant i\leqslant N-1)\big(p+q{H}_{1}+(N-2)p{H}_{1}\big).
	\notag\\	
	P_{0,j} &=\mathbbm{1}(1\leqslant j\leqslant N-1)\bigg[2\Big(1-\frac{j}{N}\Big)\Big(p{H}_{0}+p\big(1-p_{\alpha^{\text{s}}}\big)\Big)\bigg].
	\notag\\
	P_{i,i} &\!=
	\begin{cases}
	    \frac{N-2i}{N-i}\Big(p{H}_{0}+p\big(1-p_{\alpha^{\text{s}}}\big)\Big)+q{H}_{0}+q\big(1-p_{\alpha^{\text{s}}}\big),\\ \hspace{5cm} 1\leqslant i \leqslant \frac{N-1}{2}\\
		q{H}_{0}+q\big(1-p_{\alpha^{\text{s}}}\big), \hspace{0.1cm} \frac{N}{2}\leqslant i\leqslant N-1\\
		0, \hspace{0.1cm} i\geqslant N.
	\end{cases}\notag\\
	P_{1,j} &=
	\begin{cases}
		\frac{2N-2j-1}{N-1}\Big(p{H}_{0}+p\big(1-p_{\alpha^{\text{s}}}\big)\Big), \hspace{0.1cm}2\leqslant j \leqslant N-1, \\
		0,\hspace{0.1cm} j\geqslant N.
	\end{cases}\notag\\
	P_{i,1} & = 
	\begin{cases}
		\frac{2N-2i-1}{N-i}\Big(p{H}_{0}+p\big(1-p_{\alpha^{\text{s}}}\big)\Big),\hspace{0.1cm} 2\leqslant i \leqslant N-1 \\
		0,\hspace{0.1cm} i\geqslant N.
	\end{cases}
\end{align}
For $i\geqslant 2$, and $j>i$,  $P_{i,j}$ is given by
\begin{align}
	\label{pij_M1_3}
	P_{i,j} \!\!&=\!\!
	\begin{cases}
		\!\frac{N-j}{N-i}\Big(\!p{H}_{0}\!+\! p\big(1-p_{\alpha^{\text{s}}}\big)\!\Big)
		, \hspace{0.05cm}j\!+\!1\!\leqslant\! N\! \leqslant\! i\!+\!j\!-\!1\\
		\!\frac{2N-i-2j}{N-i}\Big(p{H}_{0}+p\big(1-p_{\alpha^{\text{s}}}\big)\Big),\hspace{0.05cm}N\!\geqslant\! i+j\\
		0,\hspace{0.05cm} N\leqslant j.
	\end{cases}
\end{align}
For $j\geqslant 2$, and $i>j$, $P_{i,j}$ is given by
\begin{align}
	\label{pij_M1_4}
	P_{i,j} &=
	\begin{cases}
		p{H}_{0}+p\big(1-p_{\alpha^{\text{s}}}\big),\hspace{0.1cm} i+1\leqslant N \leqslant i+j-1\\
		\frac{2N-j-2i}{N-i}\Big(p{H}_{0}+p\big(1-p_{\alpha^{\text{s}}}\big)\Big),\hspace{0.1cm} N \geqslant i+j\\
		0,\hspace{0.05cm} N\leqslant i.
	\end{cases}
\end{align}
Using the transition probabilities given in subsection \ref{NState1}, the probability that the system is in an erroneous state, $P_{E}$, or the time-averaged reconstruction error can be derived by obtaining the state stationary distributions of the transition matrix. As an example, we calculate $P_{E}$ for $N=3$
\begin{equation}
	\label{PE}
	P_{E} = \frac{{\Phi}}{\Phi+P_{2,0}-P_{2,0}P_{1,1}+P_{1,0}P_{2,1}},
\end{equation}
where $\Phi =1+P_{2,1}-P_{1,1}-P_{0,0}-P_{0,0}P_{2,1}+P_{0,0}P_{1,1}+P_{0,1}P_{2,0}-P_{0,1}P_{1,0}$.

Using \eqref{timeAvg_Error}, we define the variance of real-time reconstruction error for an observation interval $[1, T]$ as
\begin{align}
	\label{var_Error}
	\mathrm{Var}(E_t) &\!=\! \lim_{T\to\infty} \!\frac{1}{T}\sum_{t = 1}^{T} \!\mathbbm{1} (E_{t}\neq 0)^{2}\!\!-\!\!\left(\!\lim_{T\to\infty} \frac{1}{T}\sum_{t = 1}^{T} \mathbbm{1}(E_{t}\neq 0)\!\right)^{2}\notag\\& = \lim_{T\to\infty} \frac{1}{T}\sum_{t = 1}^{T} \mathbbm{1}\left(X_{t}\neq \hat{X}_{t}\right)^{2}\notag\\
	&-\left(\lim_{T\to\infty} \frac{1}{T}\sum_{t = 1}^{T}\mathbbm{1} \left(X_{t}\neq \hat{X}_{t}\right)\right)^{2}\!=P_{E}-P_{E}^{2}.
\end{align}
\begin{remark}
We can analytically prove that for a three-state DTMC information source, the randomized stationary policy has higher time-averaged reconstruction error for $p_{\alpha^{\text{s}}} < 1$ compared to the semantics-aware policy, while it has lower time-averaged reconstruction error in comparison with the change-aware policy only if $p_{\alpha^{\text{s}}}\geqslant \frac{2p}{1-p_{\text{s}}(1-2p)}$. 
\end{remark}

\section{Timing-aware Error Metrics}
\label{ConsecutiveTimeSlots}
\par In this section, we introduce two performance metrics of interest. There are several real-time and/or mission-critical applications where being consecutively in an erroneous state for some time may lead to safety issues or could even have catastrophic consequences for the system. For that, we propose the \emph{consecutive error} metric, which is defined as the number of consecutive time slots that the system is in an erroneous state\footnote{A similar metric was defined first in \cite{StamatakisGCW19} and then in \cite{AoIITON20}.}. This metric captures the temporal sequence/evolution of errors, which can cumulatively have a serious impact on the actuation performance. We also propose a companion metric, coined \emph{cost of memory error}, which considers the memory of the actuation error when the system is not in a synced state over several consecutive time slots and it can be utilized as a penalty. 

\begin{figure}[!t]
	\centering
	\begin{tikzpicture}[start chain=going left,->,>=latex,node distance=2cm,on grid,auto]
\footnotesize
\node[on chain]                        (g) {$\cdots$};
		\node[state, on chain]                 (3) {$2$};
		\node[state, on chain]                 (2) {$1$};
		\node[state, on chain]                 (1) {$0$};
		\draw[>=latex]
		(1)   edge[loop above] node {$1-p_{0,\text{e}}$}   (1)
		(1) edge  [bend left=30] node {$p_{0,\text{e}}$} (2)
		(2) edge  [bend left=30] node {$p_{\text{e},\text{e}}$} (3)
		(2) edge  [bend left=30] node[above] {$1-p_{\text{e},\text{e}}$} (1)
		(3) edge  [bend left=40] node {$1-p_{\text{e},\text{e}}$} (1)
		(3) edge  [bend left=30] node {$p_{\text{e},\text{e}}$} (g)
		(g) edge  [bend left=50] node {$1-p_{\text{e},\text{e}}$} (1)
		;
	\end{tikzpicture}
	\caption{DTMC describing the state of the consecutive error.}
	\label{Consequtive Error}
\end{figure}
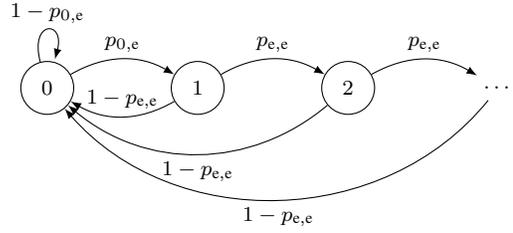

Our objective is to obtain the average consecutive error and the cost of memory error as a means to evaluate the performance. To this end, we describe the evolution of the state of consecutive error by a Markov Chain as illustrated in Fig. \ref{Consequtive Error}. In this DTMC model, the synced state is denoted by $0$, whereas $\{1, 2, \cdots\}$ denote time slots during which the system is in an erroneous state. The stationary distribution of this DTMC model can be obtained as
\begin{align}
	\label{Pn_ConsecutiveError}
	\pi_{0} = \frac{1-p_{\text{e},\text{e}}}{1+p_{0,\text{e}}-p_{\text{e},\text{e}}},\hspace{0.1cm}
	\pi_{n} = \frac{p_{0,\text{e}}(1-p_{\text{e},\text{e}})p_{\text{e},\text{e}}^{n-1}}{1+p_{0,\text{e}}-p_{\text{e},\text{e}}},
\end{align}
where $\pi_{0}$ is the probability the system is in synced state, and $\pi_{n}$ is the probability the system is in an erroneous state for $n$ consecutive time slots. Also, $p_{0,\text{e}}$, and $p_{\text{e},\text{e}}$ are defined as $p_{0,\text{e}}\! =\!\mathrm{Pr}\big[E_{t+1}\!\neq\! 0 \big|E_{t} \!= 0\big]$ and $p_{\text{e},\text{e}}=\mathrm{Pr}\big[E_{t+1} \neq 0 \big|E_{t} \neq0\big]$.

Using \eqref{Pn_ConsecutiveError}, the average consecutive error can be written as
\begin{align}
	\label{Avg_Cons_error}
	\bar{C}_{E} = \sum_{x = 1}^{\infty} x\pi_{x} = \frac{p_{0,\text{e}}}{1+p_{0,\text{e}}-2p_{\text{e},\text{e}}-p_{0,\text{e}}p_{\text{e},\text{e}}+p_{\text{e},\text{e}}^2}.
\end{align}
We define the memory of actuation error as 
\begin{align}
	\label{memoryerror}
	C_{\text{M}}(x) = 
	\begin{cases}
		0, &x = 0,\\
		\kappa^{x}, &x = 1, 2 ,\cdots, n,
	\end{cases}
\end{align}
where $n$ is finite. Using $C_{\text{M}}(x)$, we can define the cost of memory error over $n$ consecutive time slots as
\begin{align}
	\label{Avg_ConsecutiveError}
	\bar{C}^{\text{M}}_{E} \!=\! \sum_{x = 1}^{n} C_{\text{M}}(x)\pi_{x} = \frac{\kappa p_{0,\text{e}}(1-p_{\text{e},\text{e}})\big(1-(\kappa p_{\text{e},\text{e}})^n\big)}{\big(1-\kappa p_{\text{e},\text{e}}\big)\big(1+p_{0,\text{e}}-p_{\text{e},\text{e}}\big)}.
\end{align}
\vspace{-0.05in}
\section{Cost of Actuation Error}
\label{Cost_Actutation_Error}
\par In this section, we study the significance of erroneous actions at the receiver side and how different errors may have diverse impact on the system performance. For that, we consider the cost of actuation error. At time slot $t$, we denote $C_{i,j}$ the cost of error when the state of the source is $i$, i.e., $X_{t} = i$, and the state of the reconstructed source is $j\neq i$, i.e., $\hat{X}_{t}=j$. We assume that $C_{i,j}$ is fixed over time. Now, using $C_{i,j}$, we calculate the average cost of actuation error for an $N$-state DTMC model of the information source as
\begin{align}
	\label{Avg_Cost_ActuationError}
	\bar{C}_{A} = \sum_{i=0}^{N-1}\sum_{\substack{j=0 \\ j\neq i}}^{N-1} C_{i,j}\pi_{i,j},
\end{align}
where $\pi_{i,j}$ is the stationary distribution of the two-dimensional DTMC describing the joint status of the reconstructed source regarding the current state at the original source when the system is in an erroneous state, i.e., $\big(X_{t},\hat{X}_{t}\big)=\big(i,j\big), i\neq j$. In the following subsections, we consider a two- and a three-state DTMC information source and obtain $\pi_{i,j}$. Then, one can utilize the same procedure to obtain $\pi_{i,j}$ for $N$-state DTMC sources. To obtain $\pi_{i,j}$, we first assume $N=2$ and obtain the state stationary $\pi_{i,j}$ using a two-dimensional DTMC describing the joint status of the information source regarding the current state at the original source, i.e., $(X_{t}, \hat{X}_{t})$ as follows
\begin{align}
	\label{pij_M1_N2}
	\pi_{0,0}&\!\!=\!\!\frac{p\!+\!(1\!-\!p)p_{\alpha^{\text{s}}}p_{\text{s}}}{4p\!+\!2p_{\alpha^{\text{s}}}p_{\text{s}}(1-2p)}\!,\!
	\pi_{0,1}\!\!=\!\!\frac{p(1\!-\!p_{\alpha^{\text{s}}}p_{\text{s}})}{4p\!+\!2p_{\alpha^{\text{s}}}p_{\text{s}}(1-2p)}\!,\!\notag\\
	\pi_{1,0}&\!\!=\!\!\frac{p(1\!-\!p_{\alpha^{\text{s}}}p_{\text{s}})}{4p\!+\!2p_{\alpha^{\text{s}}}p_{\text{s}}(1-2p)}\!,\!
	\pi_{1,1}\!\!=\!\!\frac{p\!+\!(1\!-\!p)p_{\alpha^{\text{s}}}p_{\text{s}}}{4p\!+\!2p_{\alpha^{\text{s}}}p_{\text{s}}(1-2p)}.
\end{align}
For the change-aware policy, \eqref{pij_M1_N2} can be written as
\begin{align}
	\label{CA_M1_N2}
	\pi_{0,0} & =\pi_{1,1}= \frac{1}{4-2p_{\text{s}}},\hspace{0.1cm}
	\pi_{0,1} =\pi_{1,0}= \frac{1-p_{\text{s}}}{4-2p_{\text{s}}}.
\end{align}
Furthermore, for the semantics-aware policy, $\pi_{i,j}$ is given by
\begin{align}
	\label{SA_M1_N2}
	\pi_{0,0}&\!\!=\!\!\pi_{1,1} \!\!=\!\! \frac{p\!+\!p_{\text{s}}\!-\!pp_{\text{s}}}{4p\!+\!2p_{\text{s}}\!-\!4pp_{\text{s}}},
	\pi_{0,1}\!\!=\!\!\pi_{1,0}\!\!=\!\! \frac{p(1\!-\!p_{\text{s}})}{4p\!+\!2p_{\text{s}}\!-\!4pp_{\text{s}}}.
\end{align}
Similarly, $\pi_{i,j}$ for a three-state DTMC can be obtained as
\begin{align}
	\label{pij_M1_N3}
	\pi_{i,i} &\!=\! \frac{p\!+\!p_{\alpha^{\text{s}}}p_{\text{s}}\!-\!pp_{\alpha^{\text{s}}}p_{\text{s}}}{9p\!+\!3p_{\alpha^{\text{s}}}p_{\text{s}}\!-\!9pp_{\alpha^{\text{s}}}p_{\text{s}}}, \hspace{0.05cm} \forall i \in \{0, 1, 2\}\notag\\
	\pi_{i,j} &\!=\! \frac{p\!-\!p p_{\alpha^{\text{s}}}p_{\text{s}}}{9p\!+\!3p_{\alpha^{\text{s}}}p_{\text{s}}\!-\!9pp_{\alpha^{\text{s}}}p_{\text{s}}}, \hspace{0.05cm} \forall i,j \in \{0, 1, 2\}, i\neq j.
\end{align}
For the change-aware policy, $\pi_{i,j}$ in \eqref{pij_M1_N3} can be written as
\begin{align}
	\label{CA_M1_N3}
	\pi_{i,i}&=\frac{1+p_{\text{s}}}{9-3p_{\text{s}}},\hspace{0.2cm} \forall i \in \{0, 1, 2\}\notag\\
	\pi_{i,j}&= \frac{1-p_{\text{s}}}{9-3p_{\text{s}}}, \hspace{0.2cm}\forall i,j \in \{0, 1, 2\}, i\neq j.
\end{align}
For the semantics-aware policy, $\pi_{i,j}$ can be written as 
\begin{align}
	\label{SA_M1_N3}
	\pi_{i,i}&= \frac{p+p_{\text{s}}-pp_{\text{s}}}{9p+3p_{\text{s}}-9pp_{\text{s}}}, \hspace{0.2cm} \forall i \in \{0, 1, 2\}\notag\\
	\pi_{i,j}&= \frac{p(1-p_{\text{s}})}{9p+3p_{\text{s}}-9pp_{\text{s}}}, \hspace{0.2cm} \forall i,j \in \{0, 1, 2\}, i\neq j.
\end{align}

\section{Optimization Problem} 
\label{Optimization_Problem}
\par The objective here is to find an optimal randomized stationary policy to minimize the time-averaged reconstruction error, while keeping the time-averaged sampling cost under a given threshold. Let $\delta$ and $\delta_{\text{max}}$ be strictly positive real values, representing the cost of sampling at each attempted transmission and the total average sampling cost, respectively. Therefore, we formulate the optimization problem as
\begin{subequations}
\label{Optimization_problem1}
\begin{align}
&\underset{p_{\alpha^{\text{s}}}}{\text{minimize}}\hspace{0.3cm}\hspace{0.3cm} P_{E}\\
&\text{subject to}\hspace{0.3cm} \lim_{T \to \infty}\frac{1}{T}\sum_{t=1}^{T}\delta \mathbbm{1}\{\alpha^{\text{s}}_{t}=1\} \leqslant\delta_{\text{max}},\label{Optimization_prob1_constraint}
\end{align}
\end{subequations}
and we define $\eta = \frac{\delta_{\text{max}}}{\delta} \leqslant 1$. The constraint in \eqref{Optimization_prob1_constraint} is the time-averaged sampling cost which can be simplified as
\begin{align}
\label{Const_1}
    p_{\alpha^{\text{s}}}\leqslant \eta.
\end{align}
To solve this optimization problem we consider a two-state DTMC information source; this can be easily extended to an $N$-state DTMC with $N>2$. Using \eqref{pij_M1_N2}, the time-averaged reconstruction error can be calculated as
\begin{align}
	P_{E} = \pi_{0,1}+\pi_{1,0} = \frac{2\big(p-p p_{\alpha^{\text{s}}} p_{\text{s}}\big)}{4p+2p_{\alpha^{\text{s}}}p_{\text{s}}-4pp_{\alpha^{\text{s}}}p_{\text{s}}}\label{PE_M1_N2}.
\end{align}
Now, using \eqref{Const_1} and \eqref{PE_M1_N2}, the optimization problem can be formulated as follows
\begin{subequations}
\label{Optimization_problem1_1}
\begin{align}
&\underset{p_{\alpha^{\text{s}}}}{\text{minimize}}\hspace{0.3cm} \frac{2\big(p-p p_{\alpha^{\text{s}}} p_{\text{s}}\big)}{4p+2p_{\alpha^{\text{s}}}p_{\text{s}}-4pp_{\alpha^{\text{s}}}p_{\text{s}}}\label{PE_M1_N2_2}\\
&\text{subject to}\hspace{0.3cm} p_{\alpha^{\text{s}}}\leqslant \eta.\label{Optimization_problem1_constraint}
\end{align}
\end{subequations}
It can be readily shown that the objective function given in \eqref{PE_M1_N2_2} is decreasing with $p_{\alpha^{\text{s}}}$, that is, $\frac{\partial P_{E}}{\partial p_{\alpha^{\text{s}}}}<0$ for all values of $p$ and $p_{\text{s}}$. In other words, the objective function has its minimum value when $p_{\alpha^{\text{s}}}$ is maximum. Now, using the constraint given in \eqref{Optimization_problem1_constraint}, the optimal value of sampling probability, i.e., $p^{*}_{\alpha^{\text{s}}}$, is $\eta$. Therefore, the minimum value of the optimization problem, $P^{*}_{E}$, is obtained as
\begin{align}
    \label{PE_min}
    P^{*}_{E} =  \frac{2\big(p-p  p_{\text{s}}\eta\big)}{4p+2p_{\text{s}}\eta-4pp_{\text{s}}\eta}.
\end{align}
\begin{remark}
In what follows, RS policy is the abbreviation for randomized stationary policy, while RSC policy refers to the randomized stationary policy in the constrained optimization problem.
\end{remark}\vspace{-0.05in}
\section{Simulation Results}
\label{Simulation_Results}\vspace{-0.05in}
\par In this section, we validate our analysis and assess the performance of the sampling policies in terms of time-averaged reconstruction error and cost of memory error \footnote{In \cite{salimnejad2023real}, the performance of sampling and transmission policies is investigated in more detail with respect to a set of metrics.}. Simulation results are obtained using $10^{7}$ time slots and the parameters are set to $r = 30 \hspace{.05cm}\text{m}$, $\sigma^{2} = -100\hspace{.05cm}\text{dBm}$, $\alpha = 4$ and $P_{\text{tx}} = 1 \hspace{.05cm}\text{mW}$.
We also consider two SNR thresholds $\gamma = 0 \hspace{0.1cm}\text{dB}$ and $10\hspace{0.1cm} \text{dB}$, corresponding to $p_{\text{s}}=0.922$ and $p_{\text{s}}=0.445$, respectively.
\par The time-averaged reconstruction errors under the semantics-aware, change-aware, uniform, and RS policies for a three-state DTMC model describing the information source are shown in Table \ref{Table1} for different values of $p$, $q$, and $p_{s}$. In the uniform policy, a sample is acquired every $5$ time slots. The semantics-aware policy outperforms all other sampling policies, especially when the source is rapidly changing. \par
\begin{figure}[!t]
	\centering
\includegraphics[width=0.95\linewidth, clip]{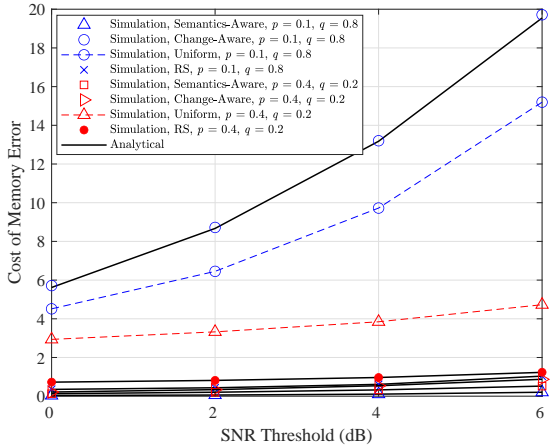}
 \caption{Cost of memory error as a function of $\gamma$, for $n = 10$, $\kappa = 2$, $p_{\alpha^{\text{s}}}=0.7$, and different values of $p$ and $q$.}
\label{Penalization_MemoryError}
\end{figure}

 In Table \ref{Table2}, we show the minimum time-averaged reconstruction error under a sampling cost constraint for $p_{\text{s}}=0.5$, $\eta=0.5$ and different values of $p$. We observe that the semantics-aware policy outperforms the optimal RSC policy for $p\leqslant\frac{\eta p_{\text{s}}}{1-2\eta+2\eta p_{\text{s}}}$ when the source is slowly varying\footnote{In the semantics-aware and the change-aware policies, the constraint of the optimization problem can be obtained as $\frac{p}{2p+p_{\text{s}}-2pp_{\text{s}}}\leqslant\eta$ and $p\leqslant \eta$, respectively.}. Note that the optimal values with red color for the semantics-aware, change-aware, and RS policies are obtained for values of $p$ and $p_{\alpha^{\text{s}}}$ that violate the constraint requirement. This means that in an unconstrained scenario, the performance of the optimal RS and the semantics-aware sampling policies is the same, however in that case, \emph{the optimal solution for the RS is to sample and transmit on every time slot, which results in the generation of an excessive amount of samples}. In sharp contrast,
a key result of this work is that when a constraint on the sampling cost is imposed, which is a practically relevant scenario, the proposed RS policy outperforms all other sampling policies for
a rapidly changing source.
Fig. \ref{Penalization_MemoryError} shows the cost of memory error for three-state DTMC information source as a function of $\gamma$ for $n = 10$, $\kappa = 2$, $p_{\alpha^{\text{s}}} = 0.7$, and different values of $p$ and $q$. As seen in this figure, as $\gamma$ increases, the cost of memory error increases. This is because when $\gamma$ increases, the success probability $p_{\text{s}}$ decreases, thus a transmitted sample is decoded with a lower probability. Note also that the semantics-aware policy exhibits smaller cost of memory error compared to all other policies. Thus, the semantics-aware policy does not allow the system to operate in an erroneous state for several consecutive time slots. 
\begin{table}[]
	\centering
	\caption{Time-averaged reconstruction error for different values of $p_{\alpha^{\text{s}}}=0.7$, $p_{\text{s}}$, $p$ and $q = 1-2p$.}
	\label{Table1}
	\begin{tabular}{|c|c|c|c|c|c|}
		\hline
		${p}$ & ${p_{\text{s}}}$ & $\text{{Semantics-aware }}$& ${\text{Change-aware}}$&${\text{Uniform}}$&${\text{RS}}$\\
		\hline
		{0.1}  & {0.922}  & {0.016} & {0.075}&{0.322}&{0.094}\\ \hline
		{0.1} & {0.445}  & {0.181}  & {0.434} &{0.485}&{0.266} \\ \hline
	{0.3}  & {0.922}  & {0.047} & {0.075}&{0.529}&{0.220}\\ \hline
		{0.3} & {0.445}  & {0.352}  & {0.434} &{0.601}&{0.443} \\ \hline
	\end{tabular}
\end{table}
\begin{table}[]
	\centering
	\caption{Minimum of reconstruction error for $p_{\text{s}}=0.5$, $\eta=0.5$ and different values of $p$.}
	\label{Table2}
	\begin{tabular}{|c|c|c|c|c|c|}
		\hline
		$p$ & $\text{Semantics-aware}$& $\text{Change-aware}$&$\text{Uniform}$&$\text{RSC}$&$\textcolor{red}{\text{RS}}$\\
		\hline
		0.1  & 0.083 & 0.333&0.299&0.187&\textcolor{red}{0.083}\\ \hline
		0.3 & 0.187  & 0.333 &0.417&0.321 &\textcolor{red}{0.187}\\ \hline
		0.5   & 0.250 & 0.333&0.450&0.375&\textcolor{red}{0.250}\\ \hline
		0.7  & \textcolor{red}{0.291}  & \textcolor{red}{0.333}&0.464&0.404&\textcolor{red}{0.291}\\ \hline
		0.9  & \textcolor{red}{0.321}  & \textcolor{red}{0.333}&0.468&0.422&\textcolor{red}{0.321} \\ \hline
	\end{tabular}
\end{table}

\vspace{-0.1in}
\section{Conclusion}\vspace{-0.05in}
\label{Conclusion}
\par We considered a time-slotted communication system where a device performs joint sampling and transmission over a wireless channel to track the evolution of a Markov source. We provided general expressions for the transition probabilities and derived the time-averaged reconstruction error, the average consecutive errors, and the cost of actuation error.
Furthermore, we formulated an optimization problem to find the optimal randomized stationary policy that minimizes the time-averaged reconstruction error under a sampling cost constraint. Our results show that the semantics-aware policy performs the best except under a sampling cost constraint and when the source is rapidly evolving, in which cases the proposed randomized stationary policy is better.
\vspace{-0.05in}
\section*{Acknowledgement}\vspace{-0.05in}
The work of M. Salimnejad and N. Pappas is supported by Zenith, the Swedish Research Council (VR), the Excellence Center at Linköping-Lund in Information Technology (ELLIIT), and the European Union (ETHER, 101096526). The work of M. Kountouris has received funding from the European Research Council (ERC) under the European Union’s Horizon 2020 research and innovation programme (Grant agreement No. 101003431).
\vspace{-0.1in}
\bibliographystyle{IEEEtran}
\bibliography{ref}
\end{document}